# Spontaneous spin-valley polarization in NbSe$_2$ at a van der Waals interface

Hideki Matsuoka[1,2], Tetsuro Habe[3,4], Yoshihiro Iwasa[1,2], Mikito Koshino[5], and Masaki Nakano[1,2,*]

[1]*Quantum-Phase Electronics Center and Department of Applied Physics, the University of Tokyo, Tokyo 113-8656, Japan.*

[2]*RIKEN Center for Emergent Matter Science (CEMS), Wako 351-0198, Japan.*

[3]*Department of Applied Physics, Hokkaido University, Hokkaido 060-0808, Japan.*

[4]*Nagamori Institute of Actuators, Kyoto University of Advanced Science, Kyoto 615-0096, Japan.*

[5]*Department of Physics, Osaka University, Osaka 560-0043, Japan.*

[*]e-mail: nakano@ap.t.u-tokyo.ac.jp

**Abstract**

A proximity effect at a van der Waals (vdW) interface enables creation of an emergent quantum electronic ground state. Here we demonstrate that an originally-superconducting two-dimensional (2D) NbSe$_2$ forms a ferromagnetic ground state with spontaneous spin polarization at a vdW interface with a 2D ferromagnet V$_5$Se$_8$. We investigated the anomalous Hall effect (AHE) of the NbSe$_2$/V$_5$Se$_8$ magnetic vdW heterostructures, and found that the sign of the AHE was reversed as the number of the V$_5$Se$_8$ layer was thinned down to the monolayer limit. Interestingly, the AHE signal of those samples was enhanced with the in-plane magnetic fields, suggesting an additional contribution to the AHE signal other than magnetization. This unusual behavior is well reproduced by band structure calculations, where the emergence of the Berry curvature along the spin-degenerate nodal lines in 2D NbSe$_2$ by the in-plane magnetization plays a key role, unveiling a unique interplay between magnetism and Zeeman-type spin-orbit interaction in a non-centrosymmetric 2D quantum material.

**Introduction**

Van der Waals (vdW) superstructures constructed with a variety of two-dimensional (2D) materials play a central role in modern condensed matter physics and materials science[1,2], providing a perspective that has never been discussed so far. A remarkable example is an emergent strong correlation effect in a moiré flat band system made of weakly-correlated materials such as graphene and semiconducting transition-metal dichalcogenides (TMDCs)[3–5], where the motion of electrons is quenched and the low-energy electronic properties are governed by the Coulombic interaction. Another intriguing example is a strong proximity effect at a vdW interface composed of atomically-thin quantum materials, where the hybridization of electron wavefunctions leads to generation of a novel quantum electronic ground state that is missing in individual materials, as exemplified by the emergent topological states observed in the graphene/TMDC heterostructures as well as in the $NbSe_2/CrBr_3$ heterostructures[6–8]. Those studies also verify a strong electronic coupling at a vdW interface despite that the constituent materials are weakly bonded via the vdW force, suggesting that a vdW interface should provide an ideal material platform for designing and creating novel physical properties and functionalities.

Among various possibilities, fabrication of a magnetic vdW heterostructure is an interesting and important research direction both for fundamental and applied researches[2,9,10]. Recent studies on the $WSe_2/CrI_3$ heterostructures have revealed that a magnetic exchange interaction is present at a vdW interface even though each layer is well separated by a vdW gap[9,10], providing the so-called "valley-Zeeman effect"[11–13] at zero field in $WSe_2$ induced by the exchange field from neighboring $CrI_3$ across the vdW interface. In this system, however, the ground state of $WSe_2$ should be still non-magnetic without spontaneous spin polarization, because $WSe_2$ is a semiconductor characterized with the fully-occupied band, where the numbers of the up-spin and down-spin electrons below the Fermi level ($E_F$) should be always equal even when proximitized by a ferromagnet. On the other hand, a magnetic proximity effect in a metallic system characterized with the partially-occupied band should lead to generation of a novel magnetic ground state with spontaneous spin polarization by producing an imbalance between the numbers of the up-spin and down-spin electrons below $E_F$. Moreover, such a metallic system should enable us to examine the low-energy electronic properties of the system by the anomalous Hall effect (AHE) measurements, through which we should be able to obtain fundamental information on the electronic structure of the system near $E_F$. Taken together, studying a magnetic proximity effect in a metallic vdW system should be of significant importance, but such an attempt has not been reported so far.

In this study, we demonstrate a proximity-induced ferromagnetic ground state in atomically-thin $NbSe_2$ at a vdW interface with a 2D ferromagnet, and uncover an intriguing feature of ferromagnetic $NbSe_2$ associated with its unique band structure by the AHE measurements. $NbSe_2$ is one of representative metallic *H*-type TMDCs showing the charge-density wave (CDW) and the superconducting (SC) transition at low temperature[14], while it is magnetically inactive due to highly delocalized nature of 4*d* electrons in $NbSe_2$. An unprecedented feature of $NbSe_2$ in the context of 2D materials research arises when thinned down to the monolayer limit, where broken in-plane inversion symmetry relevant to the trigonal prismatic structure (see Fig. 1a) and large spin-orbit interaction (SOI) lift the spin degeneracy near $E_F$. This results in generation of the out-of-plane spin-polarized electrons near $E_F$ as schematically illustrated in Fig. 1b[15–17], while the system is non-magnetic due to time-reversal symmetry. Such SOI is termed as "Zeeman-type SOI" or "Ising-type SOI", which plays an essential role in the spin-valley locking effect in monolayer *H*-type TMDCs, providing intriguing physical phenomena both in semiconducting and metallic *H*-type TMDCs[18,19]. On the other hand, an interplay between Zeeman-type SOI and magnetism has not been unveiled yet.

As for a 2D ferromagnet, we employed $V_5Se_8$, a layered magnet characterized by the 2D $VSe_2$ sheets separated by the layers of the periodically-aligned V atoms (see Fig. 1c). This compound is known to be an itinerant antiferromagnet in the bulk form[20], whereas it exhibits weak itinerant ferromagnetism with negative AHE in the thin film form, which survives down to the 2D limit[21]. We recently succeeded in constructing a magnetic vdW heterostructure with an atomically-abrupt vdW interface based on $NbSe_2$ and $V_5Se_8$ by molecular-beam epitaxy (MBE), and demonstrated that there is a strong proximity effect working at this $NbSe_2/V_5Se_8$ (Nb/V) interface characterized by the robust out-of-plane magnetic anisotropy and the large enhancement of the transition temperature[22]. Those results strongly suggest that there should be an interface exchange interaction present in the Nb/V heterostructures and that the electronic state of $NbSe_2$ should be also modulated, although no clear evidence is provided. Here, we systematically investigated the magneto-transport properties of the Nb/V heterostructures, where the number of the $V_5Se_8$ layer ($N$, varied) was set to be thinner than that of the $NbSe_2$ layer (4 L, fixed) so that the transport properties are dominated by $NbSe_2$. The details of the sample fabrication and characterization are described in our previous study[22], and brief summary is shown in Methods and Supplementary Note 1.

**Results**

**The AHE of the $NbSe_2/V_5Se_8$ van der Waals heterostructures.**

Figure 2a shows the AHE data taken at $T = 2$ K for the series of samples, where $N$ was systematically reduced from 6 L to 1.2 L. The $N = 6$ L sample exhibited negative AHE with clear magnetic hysteresis loop, indicating the robust out-of-plane magnetic anisotropy induced in $V_5Se_8$ by a strong proximity effect from neighboring $NbSe_2$ as we discussed in the previous study[22]. On the other hand, as $N$ was decreased, the AHE signal was at first suppressed at around $N$ = 3.0 L and then developed again with the opposite sign below $N$ < 3 L. Such a sign change has never been observed in the $V_5Se_8$ individual films down to the 2D limit[21], implying essential contribution from $NbSe_2$.

The evolution of the AHE signals depending on $N$ was more systematically investigated by the temperature-dependence measurements. Figure 2b shows the temperature dependence of the AH resistance at the saturated regime ($R_{AH, sat}$) for the different $N$ samples. For the thick-enough regime ($N$ = 6 L, for example), negative AHE developed monotonously below the transition temperature (~ 30 K). This behavior is basically the same as those of the $V_5Se_8$ individual films (although the transition temperatures are largely enhanced for the heterostructures as mentioned above)[21,22], suggesting that negative AHE observed with thick-enough $V_5Se_8$ should be attributed to originally-ferromagnetic $V_5Se_8$. On the other hand, as $N$ was decreased, a positive component started to develop below 25 K, and the sign of the AHE at the lowest temperature was reversed for $N$ < 3 L. Considering that the 3 L-thick $V_5Se_8$ individual film shows rather insulating behavior at low temperature (in particular below 10 K) while the 4 L-thick $NbSe_2$ individual film exhibits metallic behavior even below 2 K[21,23], it is natural to consider that the electrical conductions of the $N$ < 3 L samples at the low temperature regime are predominantly governed by $NbSe_2$ (see Supplementary Note 2). Given that the AHE is absent in a non-magnetic material, the obtained results suggest that $NbSe_2$ is in a ferromagnetic state, and positive AHE observed with thin-enough $V_5Se_8$ should be attributed to ferromagnetically-proximitized $NbSe_2$. The sign reversal of the AHE signal shown in Fig. 2a should be therefore originating from the fact that the dominant layer providing larger contribution to the electrical conduction is varied from $V_5Se_8$ to $NbSe_2$ as the number of the $V_5Se_8$ layer is reduced.

**The angle dependence of the AHE of the $NbSe_2$/$V_5Se_8$ heterostructure.**

The AHE arising from $NbSe_2$ is quite surprising in view of the fact that $NbSe_2$ is a well-known superconductor associated with CDW, where no experimental signatures of magnetism have been discussed so far. To get insights into the origin of this novel AHE in the $NbSe_2$/$V_5Se_8$ heterostructure system, we performed further experiments on the angle dependence of the AHE for those samples exhibiting positive AHE. Figure 3a displays the AHE of the $N = 2.0$ L sample taken at $T = 2$ K with different field angles ($\theta$). We here focus on the high-enough field regime, where the magnetization direction is fixed to the applied field direction. Given that the AHE signal is proportional to the out-of-plane component of the total magnetization, the signal should be reduced when the field is tilted to the in-plane direction by following $\cos(\theta)$ as schematically illustrated in Fig. 3b. However, the AHE in the present system did not follow this expected behavior. Figure 3c shows the normalized AHE signal as a function of $\theta$ taken at $\mu_0 H = 9$ T, demonstrating a substantial deviation from $\cos(\theta)$ as highlighted by the yellow-colored area. This suggests that there is an additional contribution to the AHE signal other than magnetization arising when the *in-plane* magnetization becomes finite, which has never been observed in other 2D quantum material systems. Remarkably, as will be discussed in the following sections, this intriguing angle dependence as well as the positive sign of the AHE signal could be well reproduced by theoretical calculations based on the band structure of ferromagnetically-proximitized $NbSe_2$, suggesting their intrinsic origins rather than the extrinsic origins associated with the scattering mechanisms. Moreover, such a good agreement between experiments and theory strongly suggests that $NbSe_2$ forms a ferromagnetic ground state with spontaneous spin polarization at the NbSe$_2$/V$_5$Se$_8$ vdW interface.

**Calculations of the band structure of ferromagnetic NbSe$_2$.**

Now we discuss the origin of the AHE arising from ferromagnetic NbSe$_2$. We here consider the band structure of monolayer NbSe$_2$ assuming that a magnetic proximity effect is limited to the single layer in contact with V$_5$Se$_8$, but a more realistic case with multilayer NbSe$_2$ is discussed in detail in Supplementary Note 7, which provides qualitatively the same results. Figure 4a depicts the band dispersion of monolayer NbSe$_2$ near $E_F$ along the Γ-K-M direction in the momentum space (and its time-reversal pair), showing spin splitting due to Zeeman-type SOI. Figure 4b maps the magnitude of this spin splitting energy ($\Delta_{up-down}$) in the first Brillouin Zone (BZ), which has three-fold symmetry as expected for the trigonal crystal structure. The magnitude of $\Delta_{up-down}$ at the K and K' valleys is exactly the same but opposite in sign because of time-reversal symmetry. Figure 4c illustrates the Fermi surface (FS) of monolayer NbSe$_2$ and the Berry curvature of the lower band (*i.e.*, the band having lower energy at finite **k**) in the first BZ, $\Omega_{LB}(\mathbf{k})$, both of which are symmetric against time-reversal operation. In such a case, the AH conductivity ($\sigma_{xy}$) calculated by the integration of the Berry curvature below $E_F$ over the entire BZ becomes exactly zero, and no AHE expected. Another important aspect is that the spin degeneracy is protected along the Γ-M direction due to mirror symmetry (see Fig. 1a) as shown by the dashed lines in Fig. 4b,c, which appears only at the Γ and M points in the line cut along the Γ-K-M direction (Fig. 4a).

Let us now discuss the band structure of ferromagnetic NbSe$_2$ by considering the situation when NbSe$_2$ is subjected to the out-of-plane exchange field. Figure 4d shows the band structure of monolayer NbSe$_2$ with the exchange field ($|\mathbf{M}| = 40$ meV) applied parallel to the *c*-axis (**M**//*c*). Now, the up-spin band and the down-spin band are shifted to the opposite directions due to Zeeman effect, and the magnitude of $\Delta_{\text{up-down}}$ at the K and K' valleys becomes different (see Fig. 4d). Accordingly, the FS becomes largely distorted (see Fig. 4f), and the numbers of the up-spin and down-spin electrons below $E_F$ become unequal, providing a ferromagnetic ground state with spontaneous spin polarization. In such a ferromagnetic state, we expect non-zero $\sigma_{xy}$ due to imperfect cancellation of the Berry curvature below $E_F$ between two bands.

Figure 4g shows the resultant $\sigma_{xy}$ as a function of energy calculated from the band structure of ferromagnetic NbSe$_2$, showing non-zero $\sigma_{xy}$ at $E = E_F$. Moreover, the sign of $\sigma_{xy}$ at $E = E_F$ is positive, which is consistent with the experimentally observed positive AHE for the $N < 3$ L samples, suggesting that positive AHE is indeed originating from ferromagnetic NbSe$_2$. Considering that the K and K' valleys are now energetically inequivalent, this ferromagnetic state would possibly be interpreted as a "ferrovalley" state with spontaneous valley polarization as well[24], which is a natural consequence of the spin-valley locking effect in *H*-type TMDCs associated with Zeeman-type SOI[15–17]. We note that the spin-degenerate nodal lines shown as the dashed lines in Fig. 4b,c are now moved away from the Γ-M direction, forming the closed triangular loops surrounding the K valleys (see Fig. 4e,f). The corners of those loops are contacted with the FS near the Γ valley (see Fig. 4f), which is essentially important to explain the results of the angle-dependence measurements as will be discussed in the next section.

**Calculations of the angle dependence of the AHE of ferromagnetic NbSe$_2$.**

Figure 5a presents the angle dependence of $\sigma_{xy}$ at $E = E_F$ calculated from the band structure of ferromagnetic NbSe$_2$. Surprisingly, a deviation from cos ($\theta$) that we experimentally observed for the $N < 3$ L samples (see Fig. 3c) was successfully reproduced. This characteristic deviation from cos ($\theta$) could be considered as the enhancement of the AHE signal with the in-plane fields, which could be understood as a consequence of the emergence of the additional Berry curvature with the in-plane magnetization[25], where the intersection of the up-spin band and the down-spin band (*i.e.*, the spin-degenerate nodal lines) associated with Zeeman-type SOI plays an essential role. As we explained above, those lines are originally located along the Γ-M direction without the exchange field (see Fig. 4b,c), while they are moved to surround the K valleys under the presence of the out-of-plane exchange field (see Fig. 4e,f). Importantly, the location of those nodal lines in the momentum space is determined by the magnitude of the out-of-plane magnetization ($M_z$), and $M_z$ = 40 meV satisfies the condition that the nodal lines come close to the corner of the FS near the Γ valley (see Fig. 4d,f), providing the largest deviation from cos ($\theta$) as will be discussed later.

The top panel of Fig. 5b shows the magnified view of the band structure near one of such nodal lines, corresponding to the dotted rectangular region in Fig. 4d. There is only one crossing point visible in this plot, but this is in reality distributed in the momentum space to surround the K valleys as mentioned above (also schematically shown in Fig. 5c). When the magnetization direction is tilted to the in-plane direction (for example, $\theta$ = 20º is considered in Fig. 5d), the in-plane component of the magnetization ($M_{xy}$) becomes finite, which hybridizes the up-spin band and the down-spin band and opens an energy gap along the nodal lines as shown in Fig. 5d,e. Importantly, this gap opening accompanies the generation of the Berry curvature along the nodal lines as shown in the bottom panel of Fig. 5d by the yellow-colored area. Figure 5f illustrates the distribution of this emergent Berry curvature at $\theta$ = 20º defined as $\delta\Omega_{LB}$ (**k**) = $\Omega_{LB}$ (**k**, 20º) - $\Omega_{LB}$ (**k**, 0º) in the momentum space, which exactly traces the nodal lines (loops) surrounding the K valleys. Also shown is the FS of monolayer NbSe$_2$ in a ferromagnetic state with the exchange field $M_z$ = 40 meV, which is contacted with the peak of the emergent Berry curvature $\delta\Omega_{LB}(\mathbf{k})$ near the Γ valley. In this situation, the imbalance between $\sigma_{xy}$ arising from the lower band and that from the upper band becomes maximum, providing the largest deviation from $\cos(\theta)$. We however note that such a deviation from $\cos(\theta)$ is observable in a rather broad range of the exchange field as long as $E$ is fixed at $E_F$, whereas it disappears when $E$ is increased near to the valence band top (see Supplementary Note 6). This suggests that the observed phenomena associated with the emergence of the Berry curvature with the in-plane magnetization are unique to group-V metallic $H$-type TMDCs, and that the crossing of the FS and the nodal lines near the Γ valley plays a key role for the enhancement of the AHE signal with the in-plane fields.

**Discussion**

The present study demonstrates a proximity-induced ferromagnetic ground state with spontaneous spin polarization in a metallic 2D quantum material $NbSe_2$ originally showing the CDW/SC states. We emphasize that the unique angle dependence of the AHE characterized with the enhancement of the AHE signal with the in-plane fields observed by experiments could be well reproduced by theoretical calculations based on the band structure of ferromagnetic $NbSe_2$, providing firm evidence that $NbSe_2$ forms a ferromagnetic ground state at the interface with $V_5Se_8$. Based on the comparison between the experimental and theoretical results, we estimated that the magnetic exchange interaction at the Nb/V interface should be as large as a few tens of millielectronvolt, which is larger than those reported for the $WSe_2$-based magnetic heterostructures[9,10,26] and comparable to the value recently-reported for the $WS_2$-based heterostructures[27]. As for a possible origin of the anomalous enhancement of the AHE signal with the in-plane fields, we interpret from theoretical consideration that the generation of the Berry curvature along the spin-degenerate nodal lines in 2D $NbSe_2$ by the in-plane magnetization should play an essential role, resulting from a unique interplay between magnetism and Zeeman-type SOI in 2D NbSe$_2$. Furthermore, recent demonstration of topological superconductivity in the NbSe$_2$/CrBr$_3$ heterostructures implies the possible existence of a topological superconducting state in our all-epitaxial scalable Nb/V heterostructures as well[8], which should provide an ideal platform for future topological quantum computing device applications.

**Methods**

**Sample fabrication and characterization.**

All the Nb/V heterostructures were grown on commercially-available insulating $Al_2O_3$ (sapphire) (001) substrates (SHINKOSHA Co., Ltd.) by MBE by following our previously-established process[21-23]. A substrate was cleaned by ultrasonication in acetone and ethanol, respectively, annealed in air at 1000 °C for three hours to make a surface atomically flat, and then transferred into the ultrahigh vacuum chamber with a base pressure below ~ $1 \times 10^{-7}$ Pa (EIKO Engineering, Ltd.). Prior to the film growth, a substrate surface was treated with the Se flux at 900 °C for an hour. Then, the substrate temperature was decreased down to 450 °C, and the $V_5Se_8$ layer was grown at 450 °C. During the growth, V was supplied by an electron beam evaporator with the evaporation rate of ~ 0.01 Å/s, while Se was supplied by a standard Knudsen cell throughout the growth process with the rate of ~ 2.0 Å/s. After the growth of the $V_5Se_8$ layer, the $NbSe_2$ layer was formed at the same growth temperature. Nb was supplied by an electron beam evaporator with the rate of ~ 0.01 Å/s as well. The whole growth process was monitored by reflection high energy electron diffraction (RHEED). Clear RHEED intensity oscillations were observed during the growth, confirming the 2D layer-by-layer growth mode for the formation of the $V_5Se_8$ layer and the $NbSe_2$ layer. The exact thickness was designed by counting the number of the RHEED intensity oscillation, and confirmed by the x-ray diffraction (XRD) measurement (PANalytical, X'Pert MRD) after the growth. The local structure of the obtained heterostructure was characterized by scanning transmission electron microscope (STEM) measurement (JEOL, JEM-ARM200F). All the samples were covered by insulating Se capping layers to protect their surfaces from oxidization. See Supplementary Note 1 for the details.

**Transport measurements.**

All the samples were cut into Hall-bar shape before transport measurements by mechanical scratching through Se capping layers to define the channel regions, which were typically a few hundred micrometers. The electrical transport properties were characterized by Physical Property Measurement System (Quantum Design, PPMS).

**Theoretical calculations.**

The numerical calculations were performed by using density functional theory (DFT) and a multi-orbital tight-binding model. The electronic structure of monolayer NbSe$_2$ was obtained by using quantum-ESPRESSO[28], a package of numerical codes for DFT calculations, with the projector augmented wave method. The cut-off energy was set to 50 Ry for the plane wave basis and 500 Ry for the charge density. The convergence criterion of $10^{-8}$ Ry was used in the calculations. The lattice constant was estimated to be 3.475 Å by using a lattice relaxation code in quantum-ESPRESSO.

A magnetic proximity effect on monolayer NbSe$_2$ was simulated by an exchange potential in a multi-orbital tight binding model describing electronic states in the pristine monolayer crystal. The tight-binding model was constructed on the basis of eleven Wannier orbitals, five *d*-orbitals in Nb atom and six *p*-orbitals in top and bottom Se atoms (see Fig. 1a) with spin degrees of freedom. Here, the hopping integrals were computed by using Wannier90 (ref. 29), a code to provide maximally localized Wannier orbitals and hopping integrals between them from a first-principles band structure. In this model, the exchange potential was introduced as an on-site potential represented by the Zeeman coupling $H_z = -\mathbf{M} \cdot \boldsymbol{\sigma}/2$. The direction of the magnetization was fixed to the *yz* plane with the tilting angle $\theta = \arctan(M_y/M_z)$. Here, the coupling constants $(M_y, M_z)$ are the elements of the exchange potential along the *y* and *z* axes, and assumed to be independent of orbital characters for simplicity. The Fermi energy $E_F$ was numerically estimated under the condition of charge neutrality between valence electrons and nuclei.

The Berry curvature was calculated by using a theoretical expression[30],

$$\Omega_n^z(\mathbf{k}) = -\sum_{n' \neq n} \frac{2\mathrm{Im}\langle\psi_{n\mathbf{k}}|v_x|\psi_{n'\mathbf{k}}\rangle\langle\psi_{n'\mathbf{k}}|v_y|\psi_{n\mathbf{k}}\rangle}{(\omega_{n'} - \omega_n)^2}, \qquad (1)$$

where $|\psi_{n\mathbf{k}}\rangle$ and $E_n = \hbar\omega_n$ are the wave function and the eigen-energy, respectively, of the electronic state with the band index $n$ and the wave vector $\mathbf{k}$. The anomalous Hall conductivity ($\sigma_{xy}$) was obtained by the integration of the Berry curvatures below $E_F$ in the BZ,

$$\sigma_{xy} = -\frac{e^2}{\hbar}\int_{BZ}\frac{d^2k}{(2\pi)^2}\sum_n f_n\,\Omega_n^z(\mathbf{k}), \qquad (2)$$

with the Fermi-Dirac distribution function $f_n$. The numerical calculation was performed at zero temperature.

**Data availability.**

The data within this paper are available from the corresponding author upon reasonable request.

**Acknowledgments**

We are grateful to M. S. Bahramy and K. Ishizaka for valuable discussions, and also to Y. Wang, Y. Kashiwabara, Y. Majima, Y. Tanaka, S. Yoshida, B. K. Saika, and M. Kawasaki for experimental help. This work was supported by Grants-in-Aid for Scientific Research (Grant Nos. 19H05602, 19H02593, 19H00653, 20H01840, 20H00127, and 21K13888) and A3 Foresight Program from the Japan Society for the Promotion of Science (JSPS),

and by PRESTO (Grant No. JPMJPR20AC) and CREST (Grant No. JPMJCR20T3) from Japan Science and Technology Agency.

**Author contributions**

H.M. grew and characterized the samples, performed transport measurements, and analyzed the data. T.H. and M.K. contributed to theoretical interpretation. M.N. and Y.I. supervised the study. H.M., T.H., Y.I., and M.N. wrote the manuscript. All the authors discussed the results and commented on the manuscript.

**Competing interests**

The authors declare no competing interests.

**Additional information**

**Supplementary Information** accompanies this paper at ---.

**Reprints and permission** information is available online at ---.

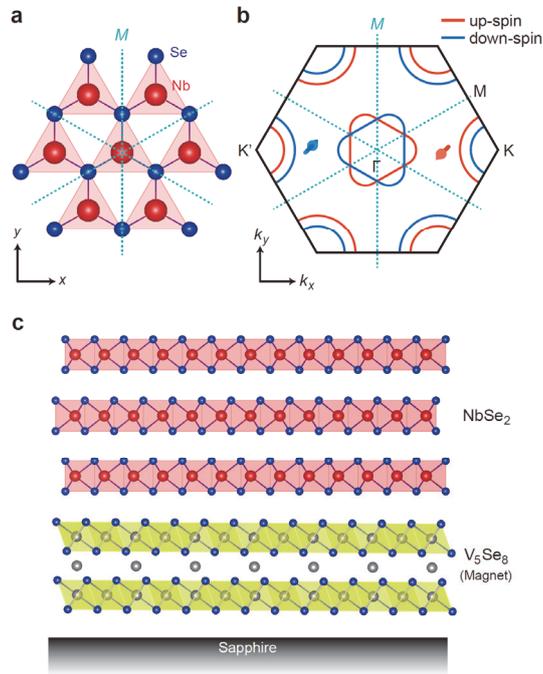

**Figure 1 | A magnetic van der Waals interface made of NbSe$_2$ and V$_5$Se$_8$.**

**a,** Schematic top view of NbSe$_2$ crystal. The dashed blue lines labeled with *M* represent the mirror planes. **b,** Schematic Fermi surface (FS) of monolayer NbSe$_2$ with Zeeman-type spin-orbit interaction (SOI). The red and blue solid lines denote the FS of the up-spin (red arrow) band and the down-spin (blue arrow) band, respectively. **c,** Schematic side view of the NbSe$_2$/V$_5$Se$_8$ (Nb/V) heterostructure.

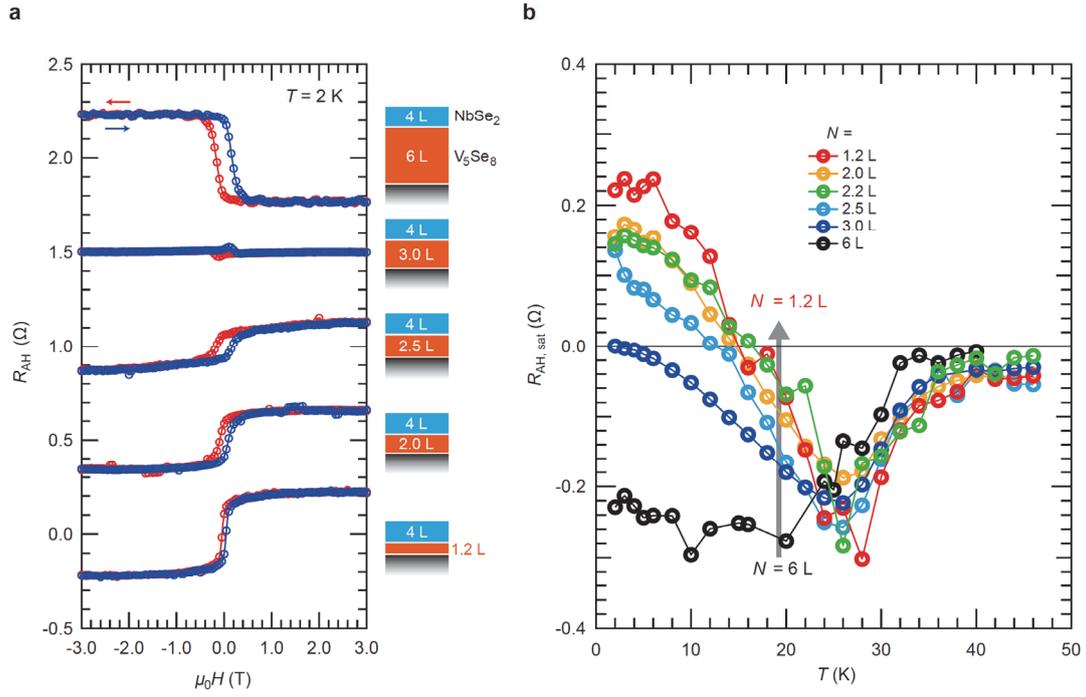

**Figure 2 | The anomalous Hall effect (AHE) of the NbSe$_2$/V$_5$Se$_8$ van der Waals heterostructures.**

**a,** The anomalous Hall resistance ($R_{AH}$) of the Nb/V heterostructures as a function of the magnetic field ($\mu_0 H$) taken at $T = 2$ K, where the number of the V$_5$Se$_8$ layer ($N$) was varied while that of the NbSe$_2$ layer was fixed to 4 L. The $N$ is defined as the number of the host VSe$_2$ layer. The data of the $N$ = 6 L, 3.0 L, 2.5 L, and 2.0 L samples are vertically shifted by 2.0 Ω, 1.5 Ω, 1.0 Ω, and 0.5 Ω, respectively. **b,** The temperature dependences of the anomalous Hall resistance at the saturated regime ($R_{AH,\,sat}$) for the different $N$ samples.

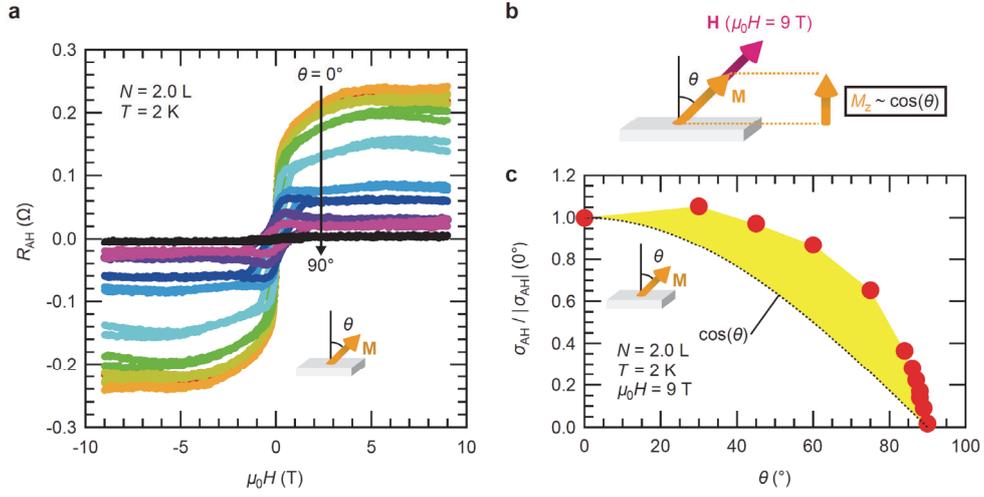

**Figure 3 | The angle dependence of the AHE of the NbSe$_2$/V$_5$Se$_8$ heterostructure.**

**a,** The AHE of the $N$ = 2.0 L sample at $T$ = 2 K with different field angles ($\theta$). The configuration of the field angle $\theta$ is shown in the inset. **b,** A schematic of the magnetization direction (**M**) and the field direction (**H**) at the high-enough field regime ($\mu_0H$ = 9 T). **c,** The AH conductivity ($\sigma_{AH}$) at $\mu_0H$ = 9 T plotted against $\theta$. The data is normalized by the $\sigma_{AH}$ at $\theta$ = 0°. The black dotted line represents cos ($\theta$) relative to the $\sigma_{AH}$ at $\theta$ = 0°, and the yellow-colored area highlights a deviation from cos ($\theta$).

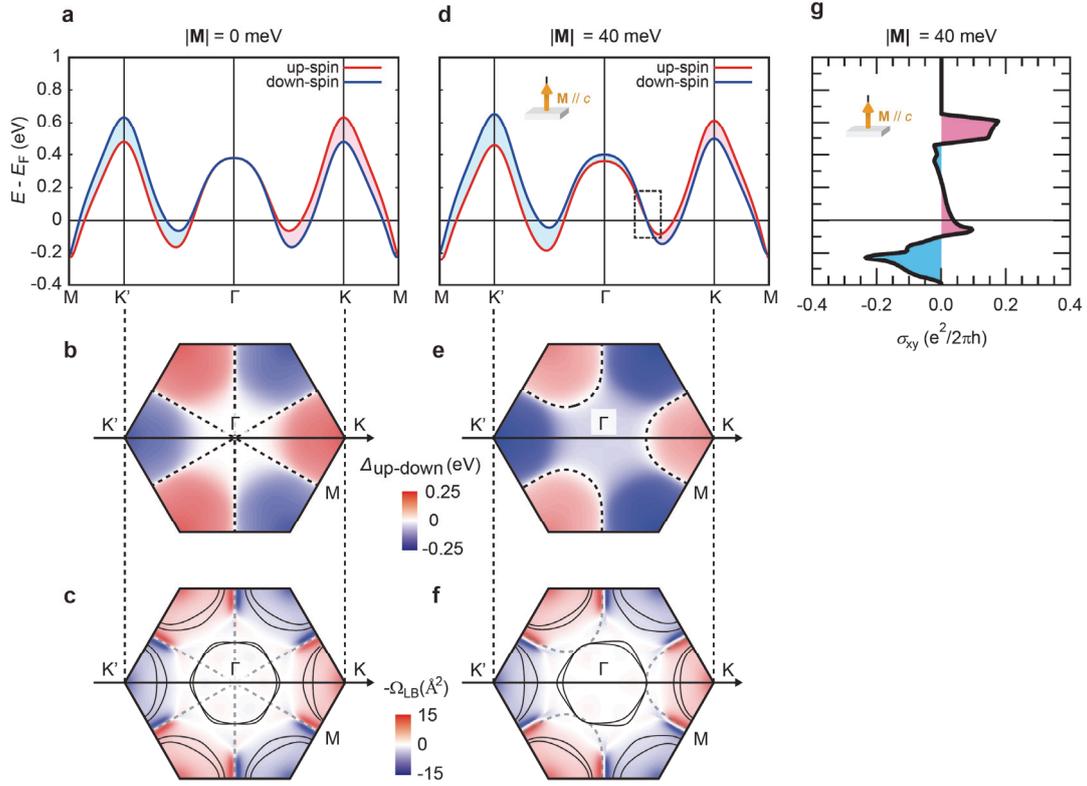

**Figure 4 | Calculations of the band structure of ferromagnetic NbSe₂.**

**a,d,** The band structure of monolayer NbSe$_2$ (**a**) without and (**d**) with the exchange field (|**M**| = 40 meV) applied parallel to the *c*-axis (**M**//*c*). The dotted rectangular region in Fig. 4d corresponds to the area highlighted in Fig. 5b,d. **b,e,** The magnitude of the spin splitting energy induced by Zeeman-type SOI ($\Delta_{up\text{-}down} = E_{up} - E_{down}$) in the first Brillouin Zone (BZ) (**b**) without and (**e**) with the exchange field. The dashed lines correspond to the spin-degenerate nodal lines. **c,f,** The FS (black solid line) and the Berry curvature of the lower band ($\Omega_{LB}$ (**k**)) in the first BZ (**c**) without and (**f**) with the exchange field. The dashed lines represent the nodal lines. **g,** The $\sigma_{xy}$ as a function of energy calculated from the band structure of monolayer NbSe$_2$ with $M_z$ = 40 meV shown in Fig. 4d.

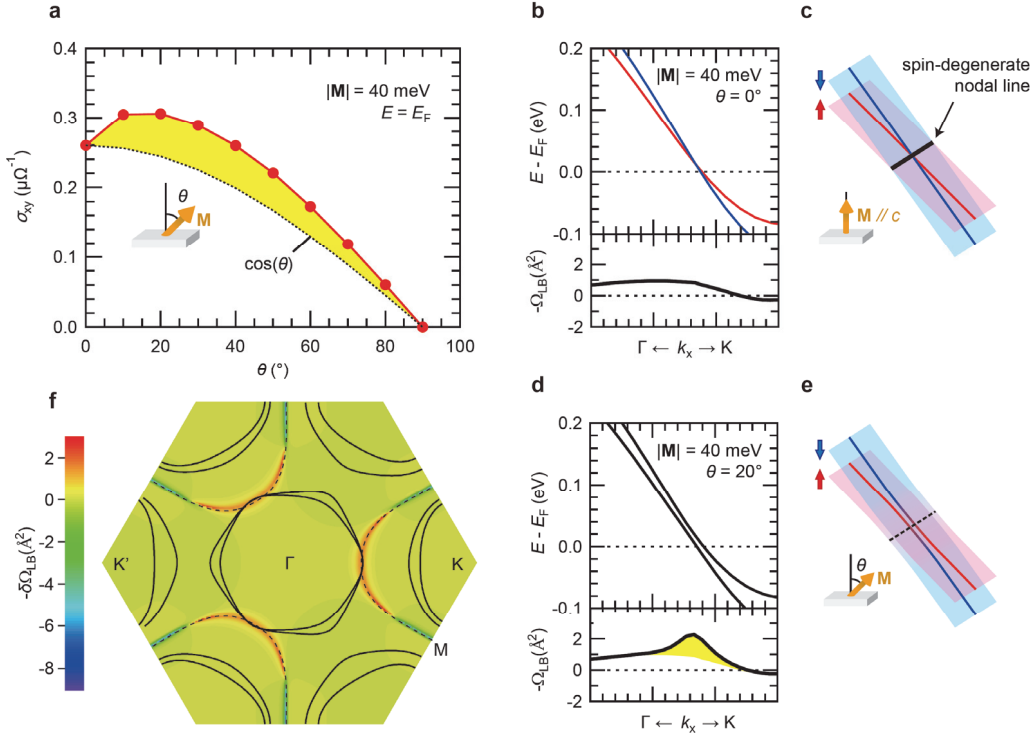

**Figure 5 | Calculations of the angle dependence of the AHE of ferromagnetic NbSe$_2$.**
**a,** The calculated angle dependence of the $\sigma_{xy}$ with $|\mathbf{M}|$ = 40 meV at $E = E_F$. The black dotted line represents cos ($\theta$) relative to the $\sigma_{xy}$ at $\theta = 0°$, and the yellow-colored area highlights a deviation from cos ($\theta$). **b,d,** The magnified views of the band structures and the Berry curvatures in the dotted rectangular region in Fig. 4d for (**b**) $\theta = 0°$ and (**d**) $\theta = 20°$. The absolute value of the exchange field is fixed to $|\mathbf{M}|$ = 40 meV. The yellow-colored area highlights the emergent Berry curvature generated by the in-plane magnetization. **c,e,** The corresponding schematics of the band structures around the spin-degenerate nodal lines. **f,** The distribution of the emergent Berry curvature at $\theta = 20°$ defined as $\delta\Omega_{LB}$ (**k**) = $\Omega_{LB}$ (**k**, 20°) - $\Omega_{LB}$ (**k**, 0°) in the momentum space. The dashed lines correspond to the nodal lines and the black solid lines illustrate the FS.

Supplementary Information for

# Spontaneous spin-valley polarization in NbSe$_2$ at a van der Waals interface

Hideki Matsuoka[1,2], Tetsuro Habe[3,4], Yoshihiro Iwasa[1,2], Mikito Koshino[5], and Masaki Nakano[1,2,*]

[1]*Quantum-Phase Electronics Center and Department of Applied Physics, the University of Tokyo, Tokyo 113-8656, Japan.*

[2]*RIKEN Center for Emergent Matter Science (CEMS), Wako 351-0198, Japan.*

[3]*Department of Applied Physics, Hokkaido University, Hokkaido 060-0808, Japan.*

[4]*Nagamori Institute of Actuators, Kyoto University of Advanced Science, Kyoto 615-0096, Japan.*

[5]*Department of Physics, Osaka University, Osaka 560-0043, Japan.*

[*]e-mail: nakano@ap.t.u-tokyo.ac.jp

**Supplementary Notes:**
1. Sample fabrication and characterization.
2. The *R-T* curves of the representative samples.
3. The detailed AHE data of the representative samples.
4. The angle dependence of the AHE of another sample.
5. The angle dependence of the normal Hall effect.
6. Other calculation results on monolayer NbSe$_2$.
7. Calculation results on bilayer NbSe$_2$.

# 1. Sample fabrication and characterization.

All the samples were fabricated by MBE by following our growth process[1-3]. The layer number was precisely designed by monitoring the RHEED intensity oscillations during the growth. Supplementary Figures 1a and 1b show typical RHEED intensity oscillations recorded during the growth of the initial $V_5Se_8$ layers and the following $NbSe_2$ layers, respectively. The actual layer numbers were confirmed by XRD measurements. Supplementary Figure 1c shows the out-of-plane XRD pattern of the $N = 6$ L sample used in this study. The strong diffraction peak was observed at around 14-15° with clear Laue oscillation, indicating high crystalline coherence along the out-of-plane direction. The complicated pattern could be well fitted by simulation calculated using the parameters of the layer number and the lattice constant for the respective layers, verifying the formation of the abrupt interface, which could be in fact confirmed by STEM measurements as shown in Supplementary Fig. 1d.

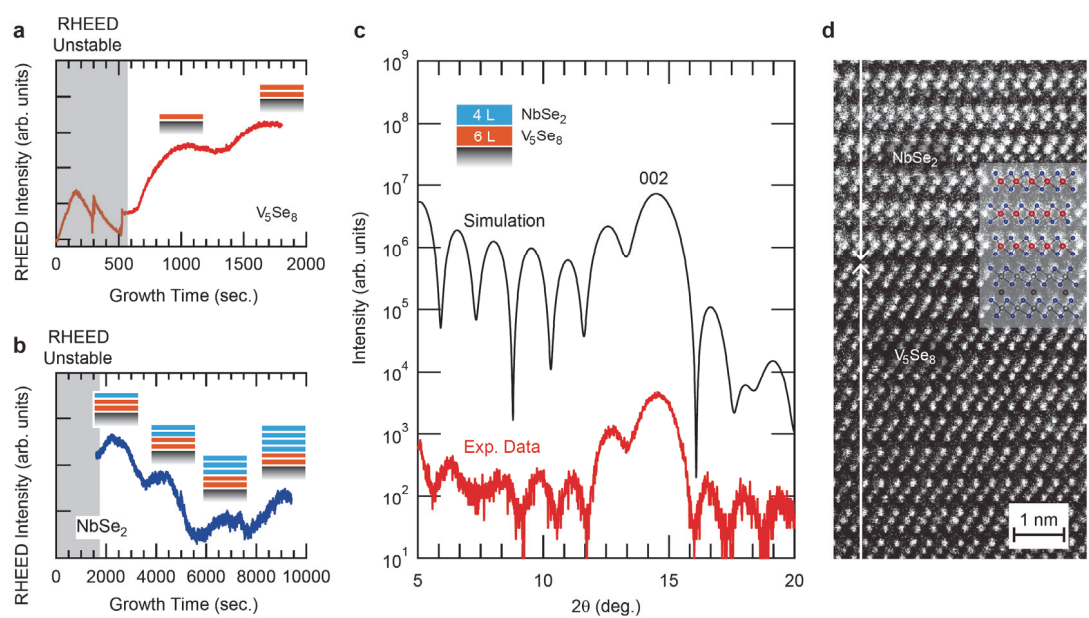

**Supplementary Figure 1 | Structural characterization of the samples.**
**a,b**, The RHEED intensity oscillations recorded during the growth of (**a**) the initial $V_5Se_8$ layers and (**b**) the following $NbSe_2$ layers. **c**, The out-of-plane XRD pattern of the $N = 6$ L sample used in this study together with the simulated pattern. **d**, The overall STEM image of a typical sample.

## 2. The *R-T* curves of the representative samples.

As we wrote in the main text, we consider that the electrical conductions of the $N < 3$ L samples in the low temperature regime are governed by the 4 L-thick $NbSe_2$ layer. Supplementary Figure 2 shows the normalized *R-T* curves of the individual films and that of the Nb/V heterostructure sample with $N = 2.0$ L. The $V_5Se_8$ individual film showed metallic behavior in the thick-enough regime (30 L), whereas it exhibited weakly insulating behavior in the thin limit (3 L) as we reported in the previous study[1]. On the other hand, the $NbSe_2$ individual film exhibited metallic behavior down to the thin limit (3 L)[2], although small upturn was still visible in the low temperature regime. The Nb/V heterostructure sample with $N = 2.0$ L exhibited metallic behavior down to the lowest temperature, whose electrical conduction should be dominated by $NbSe_2$. The characteristic kink-like behavior observed in the heterostructure sample corresponds to the ferromagnetic transition temperature[3].

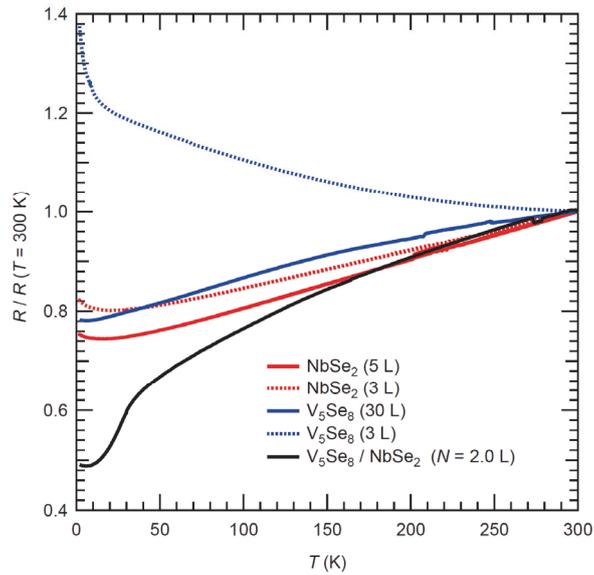

**Supplementary Figure 2 | The *R-T* curves of the representative samples.**
The normalized *R-T* curves of the 30 L and 3 L-thick $V_5Se_8$ individual films, the 5 L and 3 L-thick $NbSe_2$ individual films, and the Nb/V heterostructure with $N = 2.0$ L.

## 3. The detailed AHE data of the representative samples.

Supplementary Figures 3a-d show the anti-symmetrized AHE data of the representative samples taken at various temperatures. The data at $T = 2$ K for all four samples are the same as those shown in Fig. 2a. The temperature dependence of the $R_{AH}$ at the saturated regime, $R_{AH, sat}$, of those samples are plotted in Fig. 2b. The normal Hall components were subtracted from all the data. The sign of the AHE did not change for the $N = 6$ L sample, while it was inverted from negative to positive below $T \sim 10\text{-}20$ K for the $N < 3$ L samples.

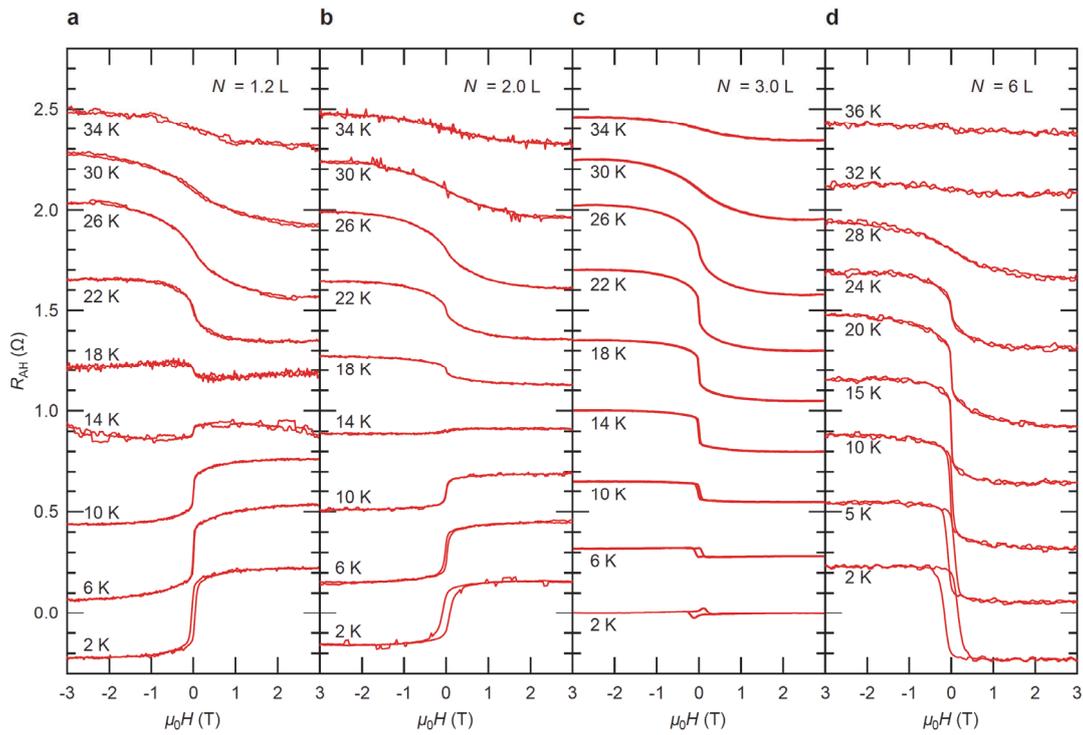

**Supplementary Figure 3 | The detailed AHE data of the representative samples.**
**a-d**, The detailed AHE data of the $N =$ (**a**) 1.2 L, (**b**) 2.0 L, (**c**) 3.0 L, and (**d**) 6 L samples, respectively, taken at various temperatures. The data at $T = 2$ K are the same as those shown in Fig. 2a. The normal Hall components were subtracted from all the data.

## 4. The angle dependence of the AHE of another sample.

Supplementary Figure 4a displays the AHE of the $N = 1.2$ L sample at $T = 2$ K with different field angles ($\theta$), and Supplementary Fig. 4b shows the normalized AHE signal as a function of $\theta$ taken at $\mu_0 H = 9$ T. The data well reproduced those for the $N = 2.0$ L sample shown in Figs. 3a and 3c.

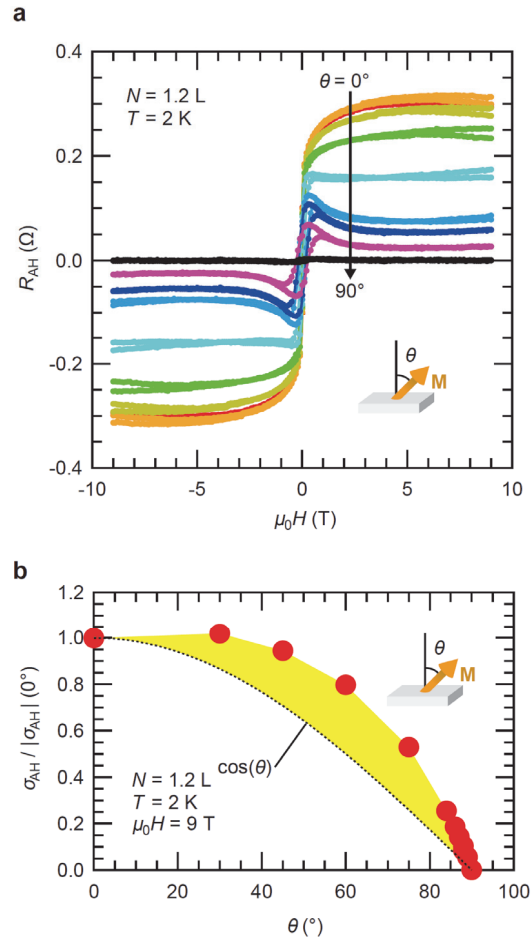

**Supplementary Figure 4 | The angle dependence of the AHE of another sample.**
**a,** The $R_{AH}$ of the $N = 1.2$ L sample at $T = 2$ K with different $\theta$. The configuration of the field angle $\theta$ is shown in the inset. **b,** The $\sigma_{AH}$ at $\mu_0 H = 9$ T plotted against $\theta$. The data is normalized by the $\sigma_{AH}$ at $\theta = 0°$. The black dotted line is cos ($\theta$) relative to the $\sigma_{AH}$ at $\theta = 0°$, and the yellow region highlights a deviation from cos ($\theta$).

## 5. The angle dependence of the normal Hall effect.

We verified the field angles ($\theta$) by checking the angle dependence of the normal Hall components. Supplementary Figures 5a and 5b show the angle dependence of the Hall coefficient ($R_H$) for (**a**) the $N = 1.2$ L sample and (**b**) the $N = 2.0$ L sample, respectively, deduced from the linear fittings of the Hall-effect data at the high-enough field regimes. In contrast to the AH components shown by the red symbols (which are the same as those shown in Supplementary Fig. 4b for $N = 1.2$ L and in Fig. 3c for $N = 2.0$ L), the normal Hall components simply follow $\cos(\theta)$ for both samples, excluding a possibility of the sample misalignment to be the origin of the deviation of the AH component from $\cos(\theta)$.

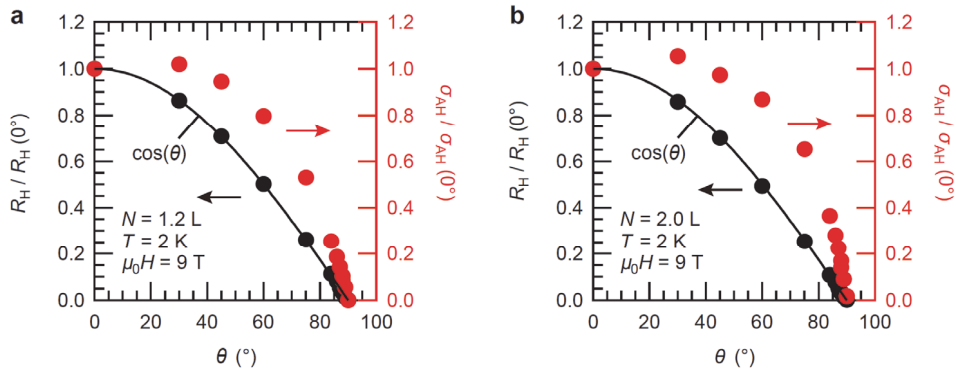

**Supplementary Figure 5 | The angle dependence of the normal Hall effect.**
The angle dependence of the normal Hall components (black symbols) and the AH components (red symbols) of the $N =$ (**a**) 1.2 L and (**b**) 2.0 L samples, respectively, taken at $T = 2$ K and $\mu_0 H = 9$ T. The data are normalized by the values at $\theta = 0°$. The black solid lines are $\cos(\theta)$ relative to the value at $\theta = 0°$.

# 6. Other calculation results on monolayer NbSe$_2$.

In the main text, we discuss the calculation results with the fixed exchange field $|\mathbf{M}| = 40$ meV, but in reality the exchange field is unknown. In this section, we provide the calculation results with different exchange fields, as well as the results on the angle dependence of the AHE at different energies.

**(I) The angle dependence of the AHE with different exchange fields.**

Supplementary Figure 6a shows the angle dependence of the $\sigma_{xy}$ with different $|\mathbf{M}|$ at $E = E_F$, and Supplementary Fig. 6b shows the magnitude of a deviation from $\cos(\theta)$ at $\theta = 20°$ as a function of $|\mathbf{M}|$. The largest deviation is achieved when $|\mathbf{M}| = 40$ meV, corresponding to the situation that the corner of the FS of monolayer NbSe$_2$ near the $\Gamma$ valley is contacted with the peak of the emergent Berry curvature surrounding the K valleys originating from the spin-degenerate nodal lines as we discussed in the main text (see Fig. 5f). We however note that such a deviation from $\cos(\theta)$ could be observed in a rather broad range of the exchange field from a few millielectronvolt to a hundred millielectronvolt (see Supplementary Fig. 6b).

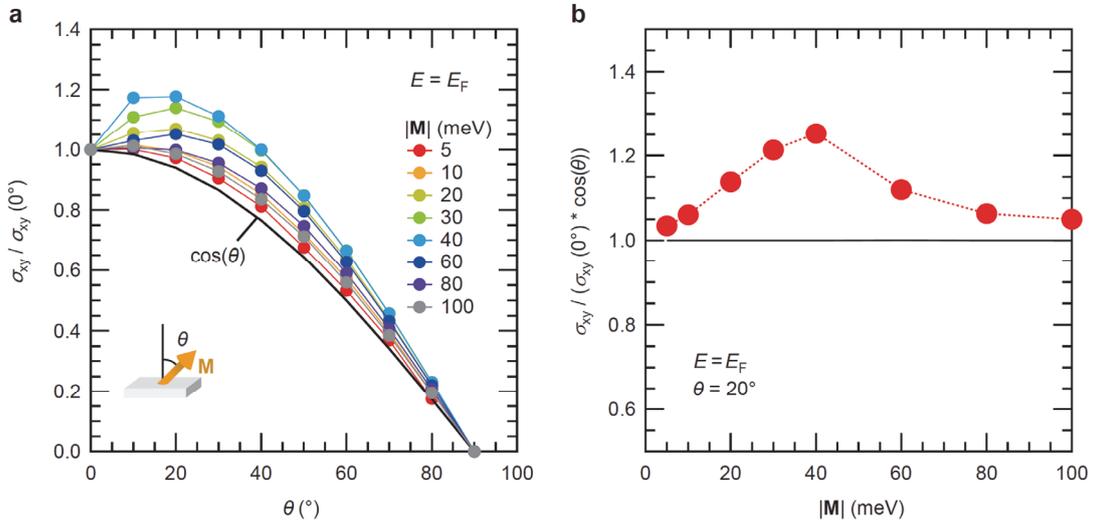

**Supplementary Figure 6 | The angle dependence of the AHE with different exchange fields.**
**a,** The angle dependence of the $\sigma_{xy}$ with different $|\mathbf{M}|$ at $E = E_F$ calculated from the band structure of monolayer NbSe$_2$. The configuration of $\theta$ is shown in the inset. **b,** The magnitude of a deviation of the $\sigma_{xy}$ from $\cos(\theta)$ at $\theta = 20°$ as a function of $|\mathbf{M}|$.

**(II) The angle dependence of the AHE at different energies.**

Supplementary Figure 7a shows the energy dependence of the $\sigma_{xy}$ with $|M| = 40$ meV for different $\theta$ calculated from the band structure of monolayer NbSe$_2$ shown in Fig. 4d, and Supplementary Fig. 7b shows the corresponding angle dependence of the $\sigma_{xy}$ at two representative energies. A deviation of the $\sigma_{xy}$ from $\cos(\theta)$ could be observed when $E = E_F$ (red symbols in **b**) but not observed when $E = E_F + 580$ meV (blue symbols in **b**). This suggests that the observed phenomena associated with the emergence of the Berry curvature with the in-plane magnetization are unique to group-V metallic $H$-type TMDCs such as NbSe$_2$ and TaS$_2$ but missing in group-VI semiconducting $H$-type TMDCs such as MoS$_2$ and WSe$_2$, and that the crossing of the FS and the nodal lines near the $\Gamma$ valley plays a key role for the enhancement of the AHE signal with the in-plane fields.

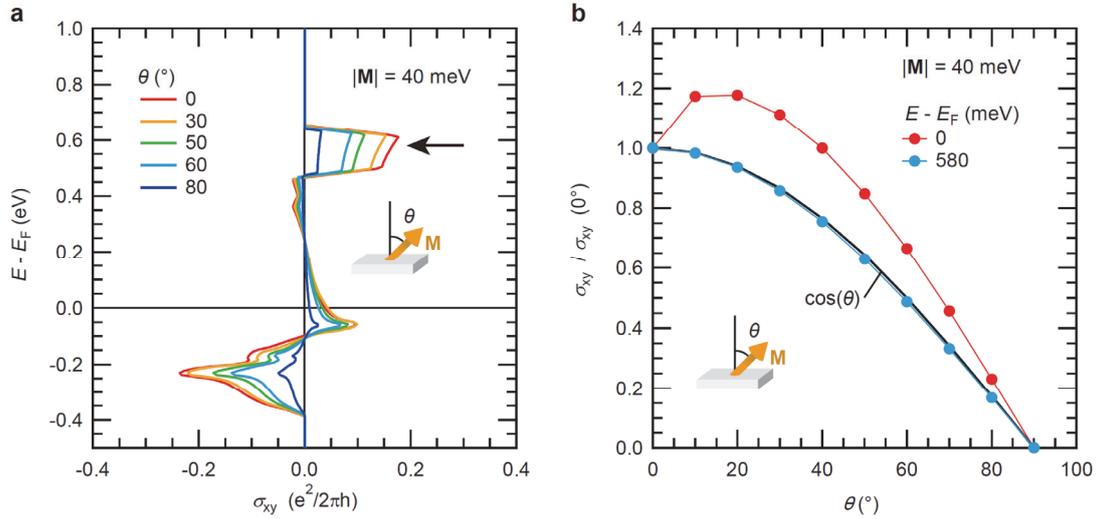

**Supplementary Figure 7 | The angle dependence of the AHE at different energies.**
**a,** The $\sigma_{xy}$ as a function of energy with $|M| = 40$ meV for different $\theta$ calculated from the band structure of monolayer NbSe$_2$. The configuration of $\theta$ is shown in the inset. **b,** The angle dependence of the $\sigma_{xy}$ with $|M| = 40$ meV at $E = E_F$ (red symbols) and at $E = E_F + 580$ meV (blue symbols), corresponding to the energy indicated by an arrow in **a**.

# 7. Calculation results on bilayer NbSe$_2$.

In the main text, we discuss the band structure of monolayer NbSe$_2$, but in reality our samples have multilayer NbSe$_2$. In this section, we discuss the band structure of bilayer NbSe$_2$ under the exchange field, and demonstrate that a physical picture based on monolayer NbSe$_2$ proposed in the main text could be applicable to bilayer NbSe$_2$ as well. Multilayer NbSe$_2$ should provide essentially the same results as those of bilayer NbSe$_2$ as long as a proximity effect is limited to one layer in contact with a ferromagnet.

**(I) The band structure of bilayer NbSe$_2$ without the exchange field.**
Supplementary Figure 8a shows a schematic crystal structure of bilayer NbSe$_2$. As compared to monolayer NbSe$_2$ with broken in-plane inversion symmetry, bilayer NbSe$_2$ has an inversion center due to the 2$H_a$ stacking, where two monolayers are stacked with 180º in-plane rotation from each other. Consequently, if we ignore the interlayer interaction, the corresponding band structure could be considered as the overlay of two monolayer bands with 180º in-plane rotation, where the up-spin/down-spin bands of the first layer (1, ↑)/(1, ↓) are fully degenerate to the down-spin/up-spin bands of the second layer (2, ↓)/(2, ↑). Then, when we consider the interlayer interaction, the bands of the different layers with the same spins get hybridized [*i.e.*, (1, ↑) hybridizes with (2, ↑), and (1, ↓) hybridizes with (2, ↓)]. The effect of this hybridization is maximum at the Γ point where the out-of-plane $d_{z^2}$ orbitals have the largest contribution[4,5], resulting in the significant modification of the bands near the Γ valley from those of monolayer NbSe$_2$.

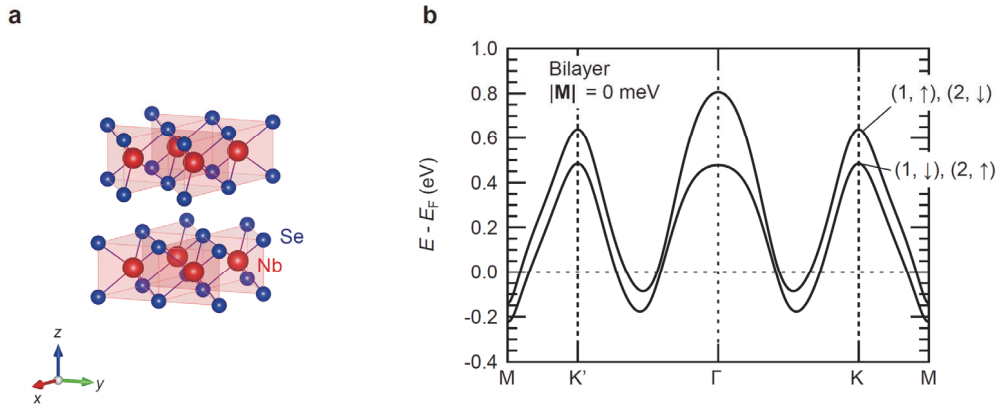

**Supplementary Figure 8 | The band structure of bilayer NbSe$_2$ without the exchange field.**
**a,** A schematic crystal structure of bilayer NbSe$_2$ with the 2$H_a$ stacking. **b,** The band structure of bilayer NbSe$_2$. The up-spin/down-spin bands of the first layer are fully degenerate to the down-spin/up-spin bands of the second layer.

On the other hand, the effect is minimum at the K and K' points where the in-plane $d_{xy}$ and $d_{x^2-y^2}$ orbitals have the largest contribution[4,5], keeping the characters of the bands near the K and K' valleys to be similar to those before hybridization. Moreover, this interlayer interaction opens an energy gap along the Γ-M line, where all the four bands are originally degenerate before hybridization. The resultant band structure is shown in Supplementary Fig. 8b. A finite energy gap is visible between the upper and lower bands, both of which have two-fold degeneracy.

**(II) The band structure of bilayer NbSe$_2$ under the exchange field.**

Now, let us consider the situation when the first layer of bilayer NbSe$_2$ is subjected to the out-of-plane exchange field. Supplementary Figure 9a shows the band structure of bilayer NbSe$_2$ with the exchange field ($|\mathbf{M}|$ = 40 meV) applied parallel to the $c$-axis, where the bands of the first layer (1, ↑)/(1, ↓) are shifted down/up due to Zeeman effect as is the case for monolayer NbSe$_2$ with the exchange field (see Fig. 4d), while the bands of the second layer (2, ↓)/(2, ↑) remain unchanged, resulting in the lifting of the degeneracy between (1, ↑)/(1, ↓) and (2, ↓)/(2, ↑). Supplementary Figure 9b shows the magnified view of the band structure around the dotted rectangular region in Supplementary Fig. 9a, where the non-degenerate four bands could be clearly recognized. Interestingly, those densely-distributed bands accompany very large Berry curvature as representatively shown for one specific band highlighted by red color on the bottom of Supplementary Fig. 9b. Such large Berry curvature in this narrow-gap region should be originating from hybridization of the orbital pseudospins by the interlayer interaction, where $d_{z^2}$ and

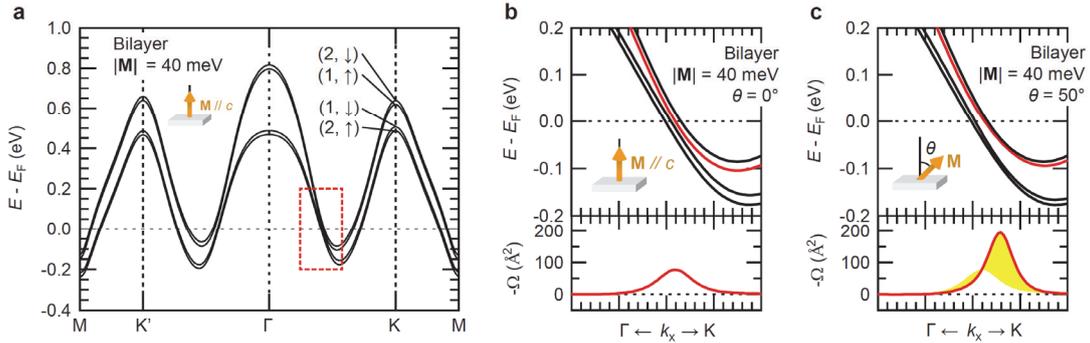

**Supplementary Figure 9 | The band structure of bilayer NbSe$_2$ under the exchange field.**

**a,** The band structure of bilayer NbSe$_2$ with the exchange field ($|\mathbf{M}|$ = 40 meV) applied parallel to the $c$-axis ($\mathbf{M}//c$). **b,c,** The magnified views of the band structures and the Berry curvature in the dotted rectangular region in **a** for (**b**) $\theta$ = 0° and (**c**) $\theta$ = 50°. The absolute value of the exchange field is fixed to $|\mathbf{M}|$ = 40 meV.

$d_{x^2-y^2} \pm i d_{xy}$ are associated with the up- and down-pseudospin, respectively[5-7]. We note that the real spins do not contribute to the Berry curvature in this case, as the up-spin band and the down-spin band are not hybridized by the out-of-plane exchange field.

Those band structure and the Berry curvature are largely modulated when the exchange field is tilted to the in-plane direction as shown in Supplementary Fig. 9c. We observe the emergence of the additional Berry curvature as is the case for monolayer NbSe$_2$, part of which should be originating from hybridization of the up-spin band and the down-spin band by the in-plane magnetization. However, there should be another contribution from the orbital pseudospins, which should be also mixed by the in-plane magnetization and generate additional Berry curvature. As a result, a deviation of the $\sigma_{xy}$ from $\cos(\theta)$ becomes much more pronounced in bilayer NbSe$_2$ as will be shown in the next section.

### (III) The angle dependence of the AHE of bilayer NbSe$_2$.

Supplementary Figure 10a shows the energy dependence of the $\sigma_{xy}$ for different $\theta$ calculated from the band structure of bilayer NbSe$_2$ shown in Supplementary Fig. 9a. The sign of $\sigma_{xy}$ is positive at $E = E_F$ as is the case for monolayer NbSe$_2$, which is consistent to the experimental results. Supplementary Figure 10b shows the angle dependences of the $\sigma_{xy}$ at $E = E_F$ for monolayer and bilayer NbSe$_2$. Clear deviations from $\cos(\theta)$ are visible for both cases, suggesting that a similar mechanism associated with the emergence of the additional Berry curvature with the in-plane magnetization should be at work for both monolayer and bilayer cases. The details of the behavior are however different, most likely because bilayer NbSe$_2$ has two origins for the additional Berry curvature, the real spins and the orbital pseudospins.

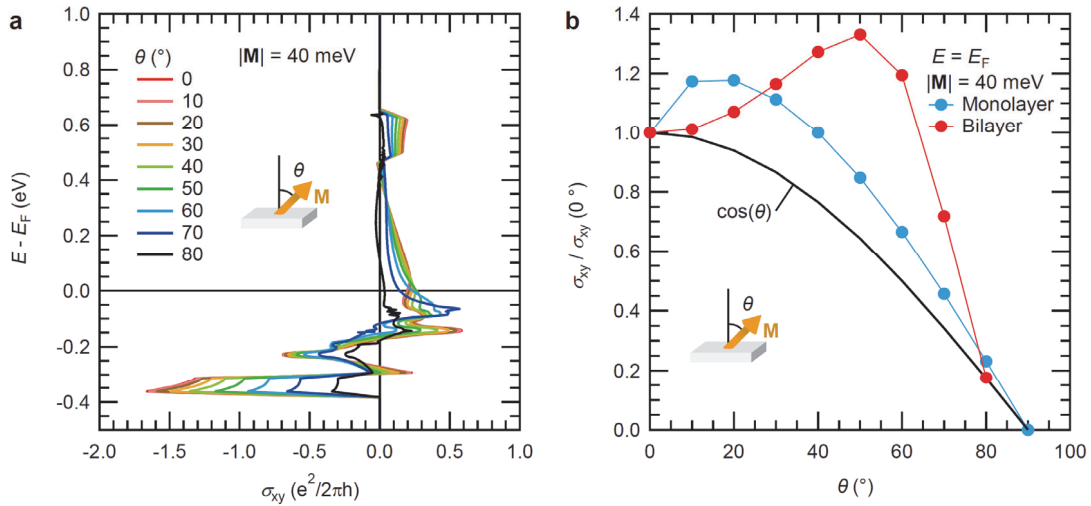

**Supplementary Figure 10 | The angle dependence of the AHE of bilayer NbSe$_2$.**
**a,** The $\sigma_{xy}$ as a function of energy with $|\mathbf{M}| = 40$ meV for different $\theta$ calculated from the band structure of bilayer NbSe$_2$. The configuration of $\theta$ is shown in the inset. **b,** The angle dependence of the $\sigma_{xy}$ with $|\mathbf{M}| = 40$ meV at $E = E_F$ for monolayer NbSe$_2$ (blue symbols) and bilayer NbSe$_2$ (red symbols).

## Supplementary References